# Martian M2 peak behavior in the dayside near-terminator ionosphere during interplanetary coronal mass ejections

Lot Ram[1], Diptiranjan Rout[2], Rahul Rathi[1], Paul Withers[3], Sumanta Sarkhel[1,*]

[*]Sumanta Sarkhel, Department of Physics, Indian Institute of Technology Roorkee, Roorkee - 247667, Uttarakhand, India (sarkhel@ph.iitr.ac.in)

[1]Department of Physics,
Indian Institute of Technology Roorkee,
Roorkee - 247667
Uttarakhand, India

[2]National Atmospheric Research Laboratory,
Gadanki, India

[3]Astronomy Department,
Boston University,
725 Commonwealth Avenue,
Boston, MA 02215
USA

## Abstract

The interplanetary coronal mass ejections (ICMEs) can pose significant impacts on the Martian ionosphere, resulting in plasma depletion, variability, and escape to space. However, the connections between the ICMEs and the associated responses of the dayside near-terminator Martian ionospheric primary peak (M2) are not well understood. The present study primarily investigates the behavior of the ionospheric peak density ($N_m$) and height ($h_m$) during the passage of ICMEs using observations from the Radio Occultation Science Experiment (ROSE) aboard MAVEN spacecraft. We have selected 8 such ICMEs (during 2017-2022) at Mars from the existing catalogs and studied the ROSE electron density profiles during quiet and disturbed time (ICMEs) for identical solar zenith angle range. We observed the elevation of the M2 peak ($h_m$ ~4-16 km) during disturbed time (ICMEs) with a decrease in $N_m$ (0.41-2.8×$10^{10}$ $m^{-3}$) in comparison to the quiet time. The present study, for the first time, addressed the influence of ICMEs on the M2 peak parameters ($N_m$ and $h_m$). We have proposed that the development of large vertical pressure gradient and electron temperature enhancement are plausible causes for ionospheric variability. Therefore, the present study provides new insights to understand peak plasma behavior in the dayside near-terminator ionosphere during ICMEs.

## 1. Introduction

The solar transient-like interplanetary coronal mass ejections (ICMEs) are the key drivers of the significant space weather events in the Martian atmosphere (Jakosky et al., 2015b; Ram et al., 2023a; Thampi et al., 2018). ICMEs are magnetized plasma structures originating from the Sun's atmosphere closed magnetic field line regions (Gopalswamy, 2006). They are accompanied by a strong rotating magnetic field, low β (ratio of plasma pressure and magnetic pressure), high proton density, enhanced dynamic pressure, and velocity (Cane and Richardson, 2003; Russell and Shinde, 2005; Zurbuchen and Richardson, 2006). Unlike Earth, Mars lacks a global intrinsic magnetic field, so the upper atmosphere, specifically the ionosphere, and localized Martian crustal magnetic field (southern hemisphere of Mars) (Acuna et al., 1998; Connerney et al., 2015b) provide a barrier to the solar wind. The solar wind-Mars interaction drives many thermal and non-thermal processes, eventually leading to depletion and escape of the Martian atmosphere over time.

The Martian upper atmosphere consists of a weakly ionized region called the ionosphere, having a peak plasma density region referred to as an M2 layer (~125-160 km) and a lower peak as an M1 layer (~105-120 km) (Cloutier et al., 1969; Bougher et al., 2017; Fox and Yeager, 2006; Withers, 2009). Previous studies showed the variability of the Martian ionospheric peak density ($N_m$) and height ($h_m$) over solar zenith angle (SZA), latitude (Lat), longitude (Lon), and lower atmospheric processes (Bougher et al., 2001; Felici et al., 2020; Forbes et al., 2002; Withers and Moore, 2020). The electron density profiles of the Martian ionosphere have been measured by various space missions in the past, viz. Mariner 4, 6-7, 9, Mars 2, 3, 4, and 6, Vikings 1 and 2, Mars Global Surveyor (MGS), and Mars Express (MEX) (Hinson et al., 1999; Kliore et al., 1972, 1973; Kolosov et al., 1972, 1973; Mendillo et al., 2006; Pätzold et al., 2004; Vasilev et al., 1975; Withers et al., 2008, 2018) at different SZA. Presently, the Mars Atmosphere and Volatile EvolutioN (MAVEN) spacecraft surveils the Martian ionosphere using in-situ and remote-sensing measurements (Jakosky et al., 2015a).

In the past, the variability in the $N_m$ has been studied in the near-terminator ionosphere with respect to SZA (Fox and Yeager, 2006; Krymskii et al., 2004; Mahajan et al., 2007; Wang and Nielsen, 2003) and solar irradiance (Chamberlain and Hunten, 1987; Schunk and Nagy, 2009; Zou et al., 2006). Additionally, both Wang & Nielsen (2003, 2004) found a rise in $h_m$ with no change in $N_m$ at fixed SZA but having a strong dependence on SZA. Also, using the photochemical model, they demonstrated that the neutral density increment during a dust storm plays the dominant role in influencing the $N_m$. Furthermore, the variation in $h_m$ as a function of longitude during dust storms has been reported by several researchers using MGS and MAVEN occultation measurements (Bougher et al., 2001; Felici et al., 2020; Girazian et al., 2019; Krymskii et al., 2003; Wang and Nielsen, 2004; Withers et al., 2018). In addition, Mahajan et al. (2009) found elevated electron density at the M2 peak during solar flares for SZA greater than 60°.

The above-mentioned studies explained the variability in $N_m$ and $h_m$ primarily based on lower atmospheric processes (e.g., dust storms), longitude, SZA, and solar flares. However, the influence of the external energy inputs in the form of ICMEs on the $N_m$ and $h_m$ of the near-terminator ionosphere has not been studied in detail motivates us to study, how the $N_m$ and $h_m$ vary with solar events like ICMEs. Also, the less-explored near-terminator region helps in understanding non-photochemical processes affecting M2 peak variation. In the present study, for the first time, we have addressed the M2 peak parameters i.e., $N_m$ and $h_m$ behavior in the dayside near-terminator ionosphere during the passage of ICMEs. For this, we utilized the MAVEN multi-instrument datasets. MAVEN in-situ instruments observe the plasma behavior at the spacecraft location with a periapsis of around 150-160 km (Jakosky et al., 2015a). In order to observe the Martian M2 peak (below 150 km), the remote-sensing measurements using the Radio Occultation Science Experiment (ROSE; Withers et al., 2020) aboard MAVEN spacecraft is utilized, which provides excellent coverage of the vertical electron density profiles. This study includes an analysis of the MAVEN ROSE-derived electron density profiles during 8 ICME events. As such, it guides more insight into the working and understanding of the ICMEs effects on the Martian peak plasma in the dayside near-terminator ionosphere. The structure of this paper is as follows: Section 2 gives a detail about the instrument datasets, Section 3 presents the observations, Section 4 provides a discussion on the variation of peak parameters, and Section 5 concludes the paper.

## 2. Data

We utilized the data from multiple instruments aboard MAVEN spacecraft for this work. The MAVEN spacecraft entered the Martian orbit in September 2014, with an apoapsis of ~6200 km and a periapsis of ~150-160 km. The upstream measurements of solar wind dynamic pressure and interplanetary magnetic field (IMF (|**B**|)) near Mars are made by Solar Wind Ion Analyzer (SWIA; Halekas et al., 2015) and Magnetometer (MAG; Connerney et al., 2015a) onboard the MAVEN spacecraft. SWIA is an electrostatic analyzer designed to measure the solar wind ions flow in the upstream region over an energy range of 5 eV-25 keV (Halekas et al., 2015). We used the MAG instrument for the magnetic field data. It measures the vector magnetic field over Mars traversed by the MAVEN orbit (Connerney et al., 2015a). The in-situ key parameter Level 2 datasets of SWIA (Halekas, 2017; Dunn, 2023) and MAG (Connerney, 2017; Dunn, 2023) are used for upstream solar wind and magnetic field observations. The SWIA and MAG datasets were accessed through Python Data Analysis and Visualization tool (PyDIVIDE; MAVEN SDC et al., 2020) and the NASA Planetary Data System (PDS).

The Martian ionospheric vertical electron density profiles are obtained using the ROSE measurements. ROSE is part of the scientific package onboard MAVEN spacecraft and works on two-way single-frequency radio occultation at X-band (7-8 GHz), a 7 GHz uplink from Earth to MAVEN and then an 8 GHz downlink from MAVEN to Earth (Withers et al., 2018). ROSE determines the vertical electron density profiles with an average vertical resolution of ~1 km and maximum uncertainty of ~3 $\times 10^9$ $m^{-3}$. The details of the radio occultation technique, measurements, data retrieval, and generation of ROSE-derived electron density processes have been provided by Withers et al. (2018), Imamura et al. (2017), and Withers and Moore (2020). The ROSE-derived Level 3 datasets (Withers, 2022) have been used to examine the vertical electron density profiles of the Martian ionosphere (~100-200 km). Also, we exploit the MAVEN datasets and CCMC (Community Coordinated Modeling Center) Space Weather Database of Notification, Knowledge, Information (kauai.ccmc.gsfc.nasa.gov/DONKI) archive to select the already identified ICMEs that could interact with Mars over the period of our study. In this study, we have selected 8 such ICMEs, where the ROSE dataset is available during both quiet and disturbed time (ICMEs).

## 3. Results

In order to investigate the ICMEs impact on the Martian peak density and height (M2 layer), we present an analysis of ROSE-derived electron density profiles during passage of 8 ICMEs from 2017-2022. These ICME events are free from dust storms and crustal magnetic fields. The following sections describe the variability of peak density ($N_m$) and height ($h_m$) in the dayside near-terminator Martian ionosphere.

### 3.1 Observations of ICMEs on Mars using SWIA and MAG instruments onboard MAVEN

Fig. 1 illustrates the variable upstream solar wind dynamic pressure and IMF conditions during ICME events. For the measurements reliably to lie in the uncontaminated solar wind intervals in the upstream region, we used the algorithm proposed by Halekas et al. (2017). The enhancement in the dynamic pressure and IMF during each ICME represents the disturbed time. The complete list of ICME events is provided in Table 1. In Fig. 1, the *x*-axis represents the days of the respective months, the *y*-axis (left, shown in green color) represents the solar wind dynamic pressure (marked by SWDP (nPa)), and the y-axis (right, shown in brown color) represents the resultant IMF (marked by |B| (nT)). During the ICMEs, the mean peak dynamic pressure and resultant magnetic field ranged from 0.73-3.53 nPa and 8.5-21.10 nT, respectively. The vertically colored dotted lines from left to right in each panel depict the time of MAVEN ROSE electron density profiles. The blue-colored dotted lines indicate the quiet time observations (before the onset of ICMEs), whereas, the red-colored lines represent the observations during the disturbed time (during ICMEs). Since MAVEN ROSE measurements occurred intermittently due to the spacecraft's elliptical orbit (Jakosky et al., 2015a) as well as operational constraints (Withers, 2020), it makes it difficult to get continuous coverage of the Martian ionosphere on the time scale. Thus, there are less numbers of profiles available in similar SZA range. In the present study, for each individual ICME event, we considered the orbit profiles (shown in blue and red colors), which lie in a similar SZA range, to compare their quiet and disturbed time behavior.

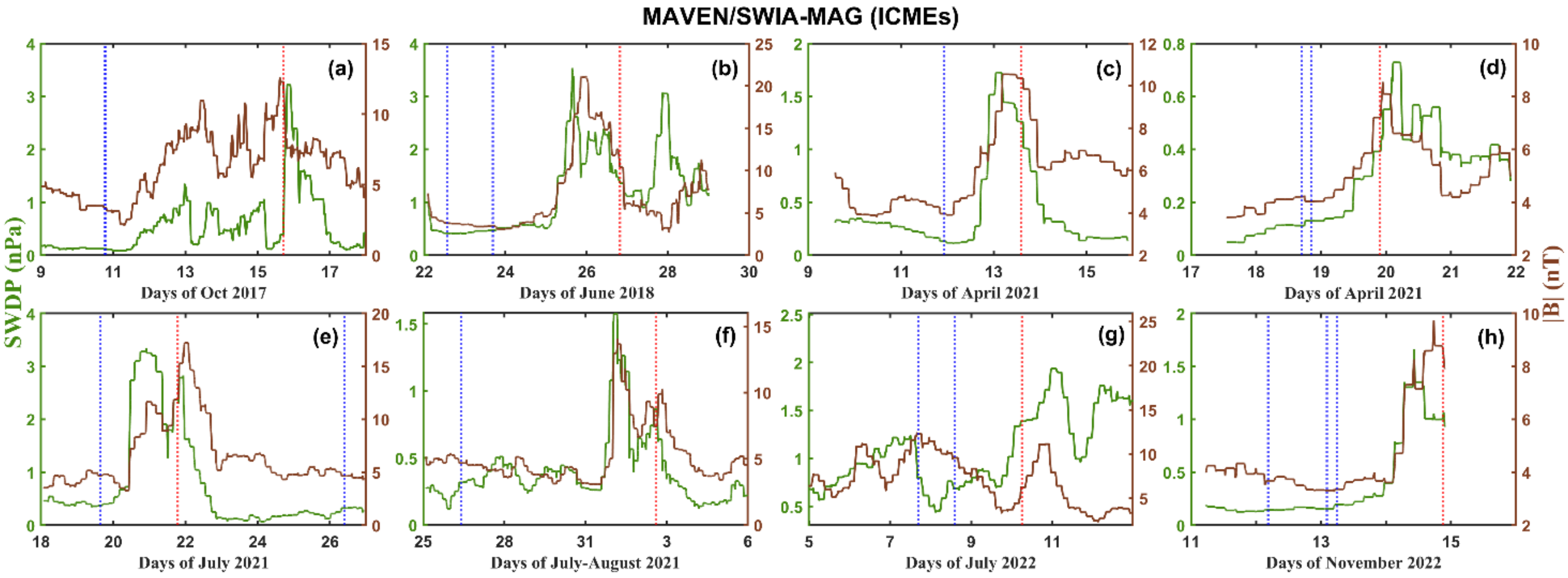

**Fig. 1**. The variations of (a-h) solar wind dynamic pressure (SWDP (nPa), left y-axis) and resultant interplanetary magnetic field (IMF |B| (nT), right y-axis) near Mars during the passage of interplanetary coronal mass ejections (ICMEs) events (2017-2022) are represented as green and brown colors scheme, respectively. The red and blue vertical-colored dotted lines indicate the observations with and without ICMEs.

### 3.2 Impact on the peak Electron Density and Altitude during the passage of ICMEs

In order to understand the ICMEs impact on the Martian dayside near-terminator ionosphere (M2 layer), we have analyzed the ROSE datasets. The ROSE measurements are marked with vertical blue and red colored dotted lines (Fig. 1) for quiet and disturbed time, respectively. A similar color scheme is used for quiet and disturbed time electron density profiles (Fig. 2). Fig. 2 shows the vertical profiles of the electron density measured by ROSE in the dayside near-terminator Martian ionosphere (SZA ~73-91°) during ICMEs. In the present study, we have considered only the dayside terminator observations because the nightside electron density profiles are highly irregular, patchy, and variable (Girazian et al., 2017; Withers et al., 2012b; Zhang et al., 1990). In Fig. 2a-h, the average quiet-time profile and 1σ standard deviation are shown with a blue color curve, which was calculated by averaging 4 km altitude bin of the profiles (at least 2-3 quiet time profiles) without ICMEs in the Martian ionosphere. Whereas, Fig. 2c & 2f show only the single quiet time density profile due to the unavailability of more density profiles at similar SZA as for disturbed time profile. The disturbed time profiles are shown in red color (Fig. 2a-h). Here, we have selected one density profile during each individual ICME, which shows a significant difference from the average quiet time profile. During each ICME, the selected quiet and disturbed time electron density profiles (blue and red; Fig. 2) are lying at similar SZA range, latitude (Lat), and longitude (Lon), so that a better comparison can be performed. Fig. 2 illustrates the variation of peak density ($N_m$) and peak height ($h_m$) during each ICME (shown in red dotted line in Fig. 1). From Fig. 2, it is evident that during the disturbed time, $N_m$ and $h_m$ show a noticeable change compared to the quiet time peak parameters. It is interesting to note that during each ICME, we observed a decrease in $N_m$ with an increase in $h_m$. During the quiet time, $N_m$ varied between $2.94\times10^{10}$ $m^{-3}$ and $8.93\times10^{10}$ $m^{-3}$, whereas during the disturbed time, they varied between $1.70\times10^{10}$ $m^{-3}$ and $7.69\times10^{10}$ $m^{-3}$. The $h_m$, during the quiet time, varied from 126-146 km, whereas for disturbed time, it varied from 142-158 km. In addition, during ICMEs, the peak altitudinal and density differences between the quiet and disturbed time vary from 4-16 km with 0.41-2.8 $\times10^{10}$ $m^{-3}$, respectively. Further, the percentage deviation in $N_m$ and $h_m$ varied between 7-42.9% and 2.74-12.70%, respectively. Table 1 provides detailed information regarding SZA, Lat, Lon, $N_m$ & $h_m$ (quiet and disturbed time), $\Delta N_m$ & $\Delta h_m$ (difference between ICMEs and Quiet Time) during each ICME event. In the present study, along with the near-terminator ionospheric region, we have also scrutinized the variability of peak density and height during other ICMEs in the dayside (48° < SZA < 60°). However, the changes in the ionospheric profiles are not prominent. The observed variabilities in the near-terminator ionospheric parameters have been investigated, and the plausible explanations are discussed in the following section.

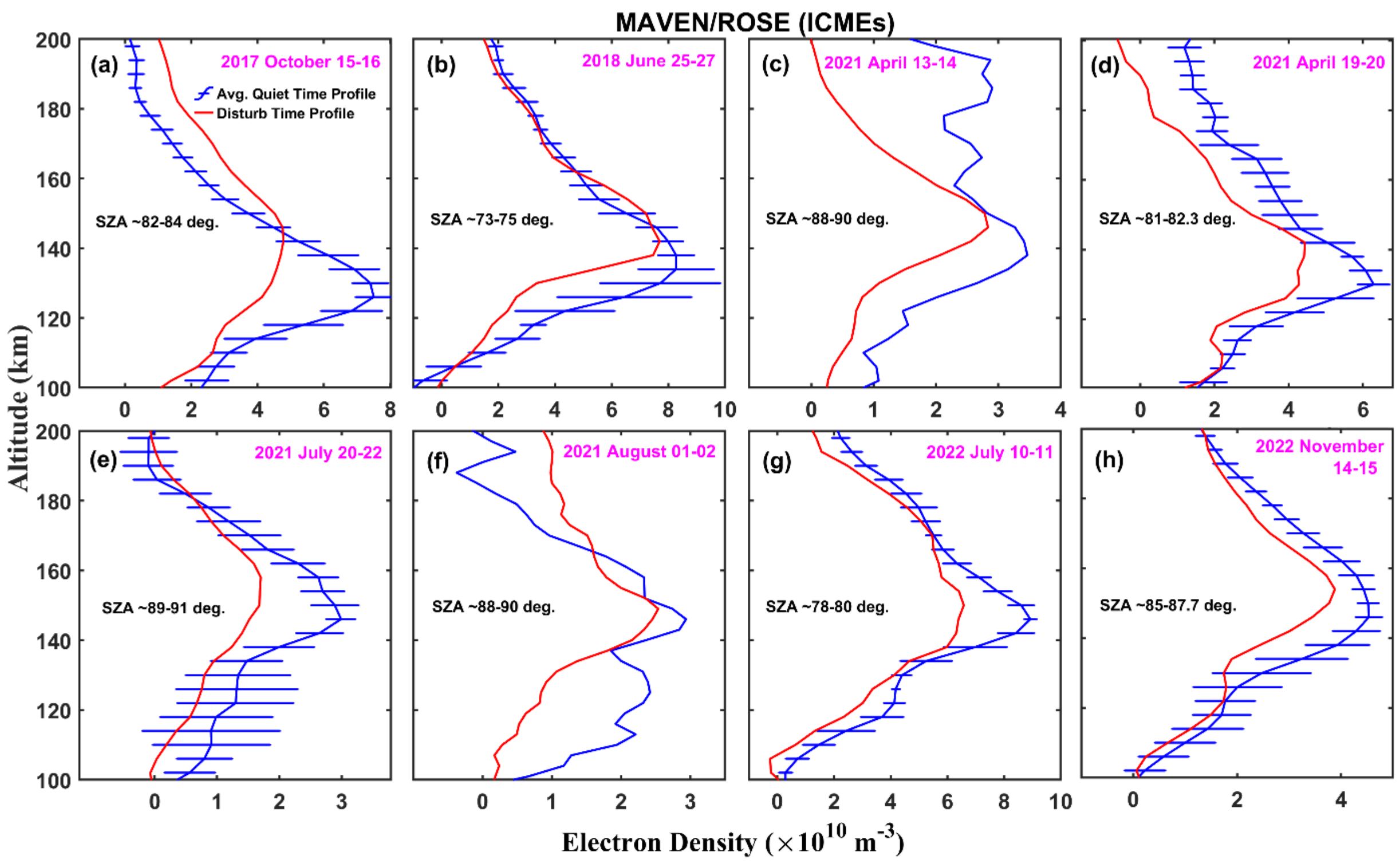

**Fig. 2.** Radio Occultation and Science Experiment (ROSE), (a-h) electron density profiles as a function of altitude (100-200 km) in the dayside near-terminator Martian ionosphere during 8 interplanetary coronal mass ejections (ICMEs) events. All profiles lie in the solar zenith angle (SZA) between 73° and 91°. The disturbed time orbit profiles are shown in red color. The average quiet time profiles are shown in blue color along with the 1σ standard deviation.

| **S. No.** | **ICMEs Start-End Date (YYYY-MM-DD)** | **SZA (°)** | **Lat (°)** | **Lon (°)** | **Peak Electron Density ($N_m$) (× 10 $^{10}$ m$^{-3}$) Quiet Time** | **Disturb Time** | **Peak Height ($h_m$) (km) Quiet Time** | **Disturb Time** | **$\Delta N_m$ (× 10$^{10}$ m$^{-3}$) Difference in Peak Density** | **$\Delta h_m$ (km) Difference in Peak Height** |
|---|---|---|---|---|---|---|---|---|---|---|
| **1.** | 2017 Oct 15-16 | 82-84 | 56-70 | 105-106 | 7.5 | 4.7 | 126 | 142 | -2.8 | 16 |
| **2.** | 2018 Jun 25-27 | 73-75 | 42-50 | 110-111 | 8.27 | 7.69 | 134 | 142 | -0.58 | 8 |
| **3.** | 2021 Apr 13-14 | 88-90 | -(84-89) | 9 & 331 | 3.46 | 2.83 | 138 | 146 | -0.63 | 8 |
| **4.** | 2021 Apr 19-20 | 81-82.3 | -(72-75) | 239-240 | 6.29 | 4.43 | 130 | 142 | -1.86 | 12 |
| **5.** | 2021 July 20-22 | 89-91 | 82-87 | 60 & 260 | 2.98 | 1.70 | 146 | 158 | -1.28 | 12 |
| **6**. | 2021 Aug 01-02 | 88-90 | 78-88 | 60 & 260 | 2.94 | 2.53 | 146 | 150 | -0.41 | 4 |
| **7.** | 2022 July 10-11 | 78-80 | 73-79 | 139-141 | 8.93 | 6.57 | 146 | 150 | -2.36 | 4 |
| **8**. | 2022 Nov 14-15 | 85-87.7 | 80-87 | 111.5-114 | 4.53 | 3.88 | 146 | 154 | -0.65 | 8 |

**Table 1.** MAVEN/ROSE observations during 8 interplanetary coronal mass ejections (ICMEs), Start-End Date, Solar Zenith Angle (SZA), Lat (Latitude), Lon (Longitude), $N_m$ (Peak Electron Density), $h_m$ (Peak height), $\Delta N_m$ (Difference in Peak Electron Density during ICMEs and Quiet time), $\Delta h_m$ (Difference in Peak Height during ICMEs and Quiet Time) during quiet and disturbed time in the dayside near-terminator Martian ionosphere.

## 4. Discussion

In the present study, we investigated the variability of the M2 layer peak density ($N_m$) and height ($h_m$) in the dayside near-terminator Martian ionosphere during 8 ICMEs. The electron density profiles (quiet and disturbed time) for each event are selected in a similar SZA range, Lat, and Lon. A noticeable change has been observed in the $N_m$ and $h_m$. The modification in ionospheric peak parameters ($N_m$ and $h_m$) during lower atmospheric processes has been reported by previous studies (Felici et al., 2020; Fox et al., 2006; Mahajan et al., 2007; Wan et al., 2022; Withers et al., 2018). It is, however, unknown whether ICMEs affect the peak density and height of the Martian ionosphere. The primary focus of the present study is to understand the M2 peak behavior in the dayside near-terminator ionosphere during the passage of ICMEs.

In the present study, the ROSE measurements show depletion in the M2 peak density and an increase in the peak height by several kilometers (~4-16 km) (Fig. 2) during each ICME in comparison to quiet time. In the earlier study by Wang and Nielsen (2004), the changes in the altitude of the peak density were observed during the Martian ionosphere interaction with the solar wind. Furthermore, due to varying solar wind conditions during ICMEs, the topside ionosphere (~200-500 km) gets more depleted, elevates ion-pick up, and escapes to space (Jakosky et al., 2015b; Ram et al., 2023a; Thampi et al., 2018, 2021) at all SZA (Girazian et al., 2019). The depletion in the topside ionosphere plausibly leads to the variability in the peak parameters ($N_m$ and $h_m$) of the ionosphere (~100-200 km), which can form a large vertical pressure gradient (Mendillo et al., 2017) between the topside and lower ionosphere. The study by Chaufray et al. (2014) also explained the steeper vertical stratification of the ionosphere than the horizontal stratification and considered vertical pressure gradient as the dominant factor. Due to the larger vertical pressure gradient, an upward movement of plasma from the lower ionosphere to the topside ionosphere, which uplifts plasma and results in the shifting of the ionospheric peak. This may lead to decrease in $N_m$ magnitude due to the redistribution of plasma from the lower to higher altitude region. This could be one of the plausible explanations for the variability of the $N_m$ and $h_m$ during the passage of ICMEs in the present study.

Furthermore, the ICMEs interaction induces horizontal magnetic fields in the Martian ionosphere (Cloutier et al., 1999) and increases the electron temperature at the ionospheric peak (Dobe et al., 1993; Krymskii et al., 2003). According to Bougher et al. (2001), the electron temperature can be controlled by solar wind interaction with varying solar wind dynamic pressure. During the passage of ICMEs at Mars, Ram et al. (2023a) and Thampi et al. (2018) observed the electron temperature enhancement above 150 km. The enhancement reinforces the vertical plasma transport to maintain the hydrostatic equilibrium, resulting in an upward drift of plasma peak height (Evans, 1971; Mendillo et al., 2018). Also, during high solar forcing like ICMEs, $\mathbf{J} \times \mathbf{B}$ acceleration (Dubinin et al.,1993a), ambipolar electric field accelerates the ion heating (Zhang et al., 2023), driving direct ion outflow, specifically $O_2^+$, exceeding that of dissociative recombination of $O_2^+$ by roughly an order of magnitude (Ergun et al., 2016). Further, it indicates that the near-terminator ionosphere is not only influenced by the photochemical equilibrium (dayside) but also by transport processes (having faster transport time than chemical lifetime; Cravens et al., 2017), which becomes more pronounced during ICMEs. Nevertheless, the observed decrease in $N_m$ and rise of $h_m$ connection in our study cannot be explained in-totality with our existing theoretical understanding. It requires complementary observations spanning both the lower and upper atmosphere for future Mars science missions. Although, there are numerous studies and models which argue that the variability in $N_m$ and $h_m$ could possibly be linked to dust storms and lower atmospheric processes (tides) (Bougher et al., 2001, 2004; Fox et al., 2006; Hanson and Mantas, 1988; Martinis et al., 2003; Rao et al., 2020; Tubiska, 2004). The study by Fang et al. (2020) showed the maximum electron density perturbation of ~5% at 150 km altitude by thermal tides using LPW. However, in the present study, the electron density perturbation of 7-42.9% has been observed during ICMEs, which is higher as compared to perturbation induced by thermal tides. In addition, the selected ROSE-derived electron density profiles in our study do not lie in the global dust storms and crustal magnetic fields. Thus, the observed variations in the M2 peak parameters in this study are free from tidal, dust storms, and crustal magnetic field effects. Furthermore, the varying solar extreme-ultraviolet (EUV) flux at Mars during solar minima (solar cycle 24/25) falls within the range of 1.5-2 $\times 10^{-4}$ $Wm^{-2}$ $nm^{-1}$ (Sánchez-Cano et al., 2023). Considering the ROSE occultation electron density profiles during 8 ICMEs under investigation have a time span of a few seconds to 5 minutes. Hence, the EUV flux variation is indeed minuscule. Further, the ROSE observations in our study are free from solar flare events (during which EUV flux rises abruptly in seconds). Thus, the impact of EUV flux is indeed minuscule. Hence, we can infer from this study that ICMEs impact is pronounced and results in the variability of the $N_m$ and $h_m$ in the dayside near-terminator ionosphere. Despite this, there is a need to further investigate the space weather effects on the terminator regions of the Martian ionosphere. This can be achieved with comprehensive datasets of the electron temperature, neutral behavior, and ion-neutral interaction in the lower and upper ionosphere.

## 5. Conclusions

In the present study, we investigated the behavior of the dayside near-terminator Martian ionosphere during the passage of ICMEs (during 2017-2022). The ionospheric peak parameters ($N_m$ and $h_m$) during quiet and disturbed (ICMEs) time have been investigated. The ROSE-derived electron density profiles at Mars during ICMEs showed an increase in $h_m$ with a decrease in $N_m$ compared to the quiet time at similar SZA. The peak height and density differences between the quiet and disturbed time vary from 4-16 km and 0.48-2.8 $\times 10^{10}$ $m^{-3}$, respectively. Also, during ICMEs, the percentage deviation in $N_m$ and $h_m$ magnitude from the quiet time varied between 7-42.9% and 2.74-12.70%, respectively. These variabilities indicate that ICMEs pose significant modification in the Martian ionosphere, which can be explained by the transport of plasma via vertical pressure gradient and electron temperature enhancement. Consequently, it results in the elevation of the ionosphere peak height with a decrement in the peak density. Our results suggest that during the passage of ICMEs, the dayside near-terminator ionospheric region is not solely controlled by photochemical equilibrium and lower processes but is also significantly affected by the transient solar events. Therefore, the present study signifies that the Martian ionosphere M2 layer is highly variable and prominently impacted during the passage of ICMEs.

## 6. Declaration of competing interest

None

## 7. Data Availability

The MAVEN datasets utilized during this work are available through the NASA PDS (https://pds-ppi.igpp.ucla.edu/search/?t=Mars&sc=MAVEN&facet=SPACECRACT

_ NAME&depth=1). The CCMC, DONKI archive is used to select the already identified ICMEs at Mars and can be accessed using https://kauai.ccmc.gsfc.nasa.gov/DONKI/. The derived data products used and produced during this work can be found in Ram et al. (2023b).

## 8. Acknowledgements

We sincerely acknowledge the MAVEN team, especially SWIA and MAG team members and NASA PDS, for the data. L. Ram acknowledges the fellowship from the Ministry of Education, Government of India, for carrying out this research work. R. Rathi acknowledges the fellowship from INSPIRE programme, DST, Government of India. This work is also supported by the Ministry of Education, Government of India.